\documentclass[twocolumn,aps,prl]{revtex4}
\usepackage{epsfig,graphicx}
\begin{document}
\title{\bf  Ab-initio density functional studies of stepped TaC
surfaces}
\author{ V.~B. Shenoy and C.~V. Ciobanu}
\affiliation{Division of Engineering, Brown University,
Providence, RI 02912 }
\date{\today}

\begin{abstract}
 We report on density functional total energy calculations of the step
formation and interaction energies for vicinal TaC(001) surfaces.
Our calculations show that double and triple-height steps are
favored over single-height steps for a given vicinal orientation,
which is in agreement with recent experimental observations.
 We provide a description of steps in terms of atomic displacements
and charge localization and predict an experimentally observable
rumpled structure of the step-edges, where the Ta atoms undergo
larger displacements compared to the C atoms.
\end{abstract}
 \maketitle

%
%
The stepped surfaces of transition metal carbides (TiC, TaC, etc.)
are technologically important due to their applications as surface
catalysts with high corrosion resistance. From a fundamental point
of view, these surfaces have received attention as they provide a
rich ground for studying faceting transitions that involve single
and multiple-height steps. In a series of experiments, Zuo and
coworkers \cite{Zuo310}--\cite{Zuo910} have investigated the
structure of several surfaces vicinal to TaC(001) and found that
multiple-height steps are favored over single-height steps in all
the cases. Motivated by these experiments, we perform ab-initio
density functional  calculations to obtain the first quantitative
results for the step formation and interaction energies for steps
of different heights on these surfaces.

The preference for single or multiple-height steps for a certain
vicinal orientation is determined by the energy required to create
an isolated step of a certain height and the repulsive
interactions between these steps. In general, it costs more energy
to create multiple-height steps, as the number of bonds that are
broken in the process of creating a step increases with its
height. On the other hand, the energetic contribution from the
repulsive step-step interactions  can be the lower in the case of
a vicinal surface with multiple-height steps due to a lowering of
the number of step pairs per unit length. As we will see, both of
these effects are borne out by our calculations. Furthermore, we
will show that the increase in the formation energies of double
and triple-height steps over the single-height steps is not large
enough to offset the effect of step-step interactions.
Consequently, the former types of steps are favored for the
vicinal orientations that we consider. In what follows, we also
will report on the structures of different types of steps in terms
of charge density and atomic displacements, and develop a
phenomenological model for the electrostatic step interactions.


The density functional calculations were carried out with the VASP
package \cite{Vasp},  using the projector augmented-wave
pseudopotentials \cite{paw} and the Perdew-Wang functional form
\cite{pw91} for exchange correlation energy. Bulk calculations
with an 8-atom cell and a 35 k-point Brillouin zone
sampling gave an optimum Ta-C separation of $a =2.2415$\AA.   
For surface calculations, the Brillouin zone was sampled using a
$\Gamma$-centered $8\times 8 \times 1 $ grid. This sampling
yielded 15 k-points for TaC(001) and 21 to 25 k-points for the
vicinal surfaces. The ions were relaxed using a conjugate-gradient
algorithm until the total energy converged to $0.01$eV. The energy
cut-off for the plane waves was set to 400eV (29.40 Ry) in all the
calculations.


Before we proceed with the discussion of vicinal surfaces, we
compare the predictions of the surface structure of TaC(001)
obtained using the present model with experimental observations
and previous theoretical studies.  It is well-known
\cite{Gruzalski}--\cite{Kobayashi} that
 this surface exhibits a rumpling reconstruction, where the C atoms
are displaced out of the surface while the Ta atoms are pulled
inwards. It can be seen from Table~\ref{flatdispl} that our
results for the relative displacements of the atoms in the first
and second surface layers are in close agreement with available
experimental \cite{Gruzalski} and theoretical \cite{Kobayashi}
data.
\begin{table}
\begin{tabular}{l |cccccc}
\hline  \hline
  & $d_C ^1$ & $d_{Ta}^1$ & $d_C^2$ & $d_{Ta}^2$ & $r_1$ & $r_2$
  \\ \hline
Exp.   \cite{Gruzalski}   & 0.09  & -0.11  & 0.04 & 0.00 & 0.20 & 0.04 \\
Theor. \cite{Kobayashi}  & 0.097 & -0.132 &   --  & --   & 0.23 & 0.052 \\
This work            & 0.077 & -0.115 & 0.027 & -0.008 & 0.192 & 0.035  \\
\hline \hline
\end{tabular}
\caption{Rumpling relaxation of TaC(001) compared with previous
results \cite{Gruzalski}, \cite{Kobayashi}. The displacements of
the first ($d_{C, Ta}^1$) and second($d^2_{C,Ta}$) layer surface
atoms, and the rumpling amplitudes $r_1$, $r_2$ (defined in
\cite{Gruzalski}) are given in \AA .} \label{flatdispl}
\end{table}

\begin{figure}
  \begin{center}
  \includegraphics[width=3.7in]{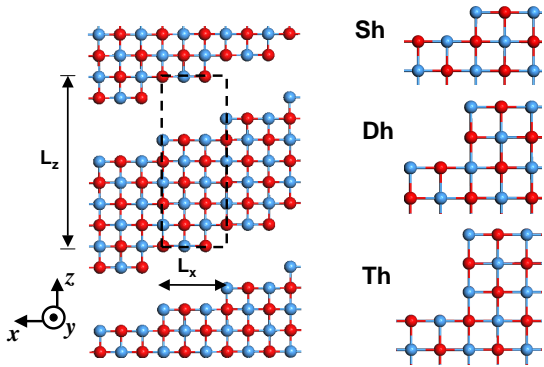}
  \end{center}
  \vspace{-7cm}
\caption{Geometry of a typical stepped surface, with the supercell
indicated by dashed lines. The structures of the single (Sh),
double (Dh) and triple (Th) -height steps considered in the
present work are also given in the figure.} \label{steptypes}
\end{figure}

%
%
For vicinal surfaces, we used supercells as illustrated in
Fig.~\ref{steptypes}, where $L_x$  and $L_z$ denote the
terrace-width of the stepped surface and height of the supercell,
respectively. The dimension of the supercell in the $y$-direction
($L_y$) is determined by the period in the direction parallel to
the step, which in the present case is the lattice constant of
TaC, $2a$. To create the steps, we employ shifted (i.e.
non-orthogonal) boundary conditions \cite{Poon}, in which the
amount of shift in the $z$- direction determines the height of the
steps. The structures of the single (Sh), double (Dh) and triple
(Th) -height steps  that we consider in the present work are shown
in Fig.~\ref{steptypes}. All calculations on vicinal surfaces were
carried out using a slab of TaC with 8 atomic rows (17.932 \AA)
and a vacuum thickness of 10\AA. In order to ensure that these
results are not influenced by finite-size effects, we repeated the
calculations for certain selected step structures and terrace
widths by increasing the cell thickness to 10 atomic rows and the
dimension of the vacuum to 15\AA; the results were found to change
by less than 5\% in all the cases that were considered.

\begin{table}
\begin{tabular}{l c c c c l}
\hline \hline
$n$   &  $\lambda _n^f$(meV/\AA) &    $ \lambda_n^d $(meV\AA)    & $ \beta_n^f $(meV/\AA$^2$)  & $ \beta_n^d $(meV/\AA$^2$) \\
\hline

1 & 268.89  &  998.42  & 119.96 & 88.85 \\
2 & 466.47  &  550.34  & 104.05 & 6.11  \\
3 & 637.08  &  1217.92 & 94.74  & 4.01  \\
\hline \hline
\end{tabular}
\caption{Step formation energies $\lambda_n^f$ and repulsion
strengths $\lambda_n^{d}$ obtained using linear fits of data in
Fig.~\ref{ledgefits} to Eq.~(\ref{lambdalin}). The table also
shows the scaled step formation energies $ \beta_n^f $ and
interaction strengths $\beta_n^d $ introduced in
Eq.~(\ref{anisose}). }\label{betas}
\end{table}

%
%

\begin{figure}
  \begin{center}
   \includegraphics[width=3.7in]{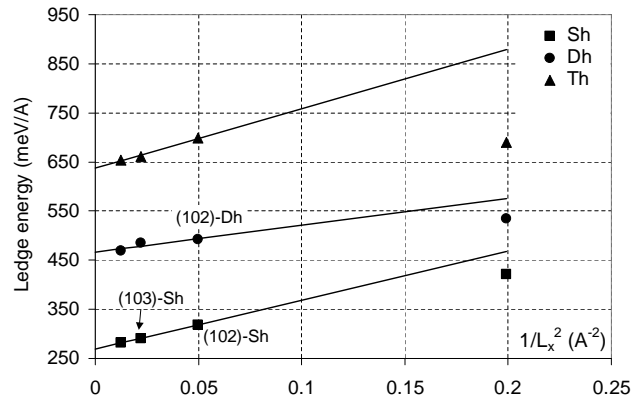}
  \end{center}
   \vspace{-7.cm}
\caption{Ledge energies $\lambda$ for the single (Sh), double (Dh)
and triple (Th) height steps plotted as a function of $1/L_x^2$.
The points for which the step separations exceed $a=$2.24\AA \ are
fit to Eqn.~(\ref{lambdalin}), $\lambda_n=
\lambda_n^f+\lambda_n^d/L_x^2$ \cite{later}. Parameters
$\lambda_n^f$, $\lambda_n^d$ are given in Table~\ref{betas}.}
\label{ledgefits}
\end{figure}

The energetics of stepped surfaces can be understood by
considering the ledge energy,  defined as $\lambda_n=( E - N_p
e_b-\gamma_{0}A )/2L_y$ \cite{Poon},  where $E$ is the total
energy of the $N_{p}$  Ta-C pairs in the slab, $e_b = -22.1536$eV
is the bulk energy per pair, $\gamma_0 = 95.53$meV/\AA$^2$ is the
surface energy of the TaC(001) surface and the subscript $n$
denotes the step-height. The ledge energy is the energy per unit
length of the vicinal surface in excess of the surface energy of
the terraces that separate the steps, and includes both the step
formation and interaction energies.  Using the fact that both the
elastic and electrostatic effects give rise to dipolar
interactions \cite{marchenko,Jay}, the ledge energy can be
expressed as
\begin{equation}
    \lambda_n = \lambda_{n}^f +\frac{\lambda_{n}^d}{L_x^2} \ \ \ ,
    \label{lambdalin}
\end{equation}
where $\lambda_{n}^f$ and $\lambda_{n}^d$ denote the formation
energy and the dipolar interaction strength of a step with height
$na$, respectively. The computed ledge energies for the three
types of steps are plotted in Fig.~\ref{ledgefits}, along with
linear fits to the functional form assumed in Eq.
(\ref{lambdalin}) \cite{later}. The fitting parameters
$\lambda_{n}^f$ and $\lambda_{n}^d$ are given in
Table~\ref{betas}.

\begin{table}
\begin{tabular}{ c c c c}
\hline \hline
$(hkl)$  \ \  & step \  \ & $\gamma$ (from data in Fig.~\ref{ledgefits}) \ \ & $\gamma$ (from Eq.~(\ref{anisose}))\\
\hline
(102)   &   Sh  & 149.01 & 146.71 \\
        &   Dh  & 132.22 & 131.67 \\
        &   Th  &  --      & 127.46 \\ \hline

(103)   &  Sh & 131.66   & 131.66 \\
        &  Dh &     --     & 123.74 \\
        &  Th &  --        &  120.73 \\
\hline
 \hline
\end{tabular}
\caption{Surface energies (in meV/\AA$^2$) of the (102) and (103)
surfaces considered by Zuo and coworkers in \cite{Zuostep}. The
table gives surface energies calculated using Eq.~(\ref{anisose})
and the scaled fitting parameters in Table~\ref{flatdispl}, as
well as surface energies for (102)-Sh and Dh and (103)-Sh
calculated directly from the ledge energies marked in
Fig.~\ref{ledgefits}. The close agreement between the surface
energies obtained using the two different approaches indicates
that Eq.~(\ref{lambdalin}) gives an accurate description of the
ledge energies.}
 \label{surfene}
\end{table}


It can be seen from Table~\ref{betas} that the formation energies
of the steps increase with their heights, in agreement with bond
counting arguments presented earlier. The dipolar interactions, on
the other hand, do not show a monotonic trend; the interaction
strength of the Dh step is smaller than the corresponding values
for the Sh and Th steps. The relative importance of these two
contributions can be understood by considering the surface energy
of a vicinal surface made of steps with height $na$, given by
\begin{equation}
\gamma_n(\theta) = \gamma_0\cos\theta + \beta_n^f |\sin \theta| +
\beta_n^d \frac {|\sin^3 \theta|}{\cos^2 \theta}, \label{anisose}
\end{equation}
where $\theta$ is the vicinal angle and the parameters $\beta_n^f
= \lambda_n^f/(na)$ and $\beta_n^d = \lambda_n^d/(na)^3$ denote
scaled step formation and interaction contributions, respectively.
Using the values of these parameters tabulated in
Table~\ref{betas}, we find that both the scaled formation and
interaction parts show a monotonic decrease with increasing step
height, indicating that multiple-step heights are favored over the
single height steps for any vicinal orientation. In the following
paragraph we consider experimental work of Zuo and coworkers
\cite{Zuo310}--\cite{Zuo910} and show that the predictions of our
calculations are in agreement with their observations.

Using a statistical analysis \cite{einstein} of the step
separation distributions on TaC(103) and (102) surfaces, Zuo and
coworkers have determined that the former surface consists of a
large number of triple-height steps and a smaller number of
double-height steps while the latter surface consists of almost
equal amounts of areas that contain either triple or
quadruple-height steps. In order to understand these results, in
Table~\ref{surfene}, we have used Eq.~(\ref{anisose}) and the data
in Table~\ref{betas} to calculate the surface energies of these
vicinal surfaces made up of Sh, Dh and Th steps. For both the
surfaces considered in the experimental work, we find that the
single-step vicinal is much larger (by $\sim$ 10-20 meV/\AA$^2$)
in energy than the vicinals with multiple-height steps.
Furthermore, the triple-height steps are favored over the
double-height steps (by $\sim$ 4 meV/\AA$^2$) in both the cases,
explaining their stability and their presence on (103) and (102)
surfaces. The presence of a few Dh steps and the complete absence
of Sh steps on (103) surfaces is consistent with the fact that the
former is closer in energy to the Th steps than the latter by
about 8 meV/\AA$^2$ (refer to Table~\ref{surfene}). At the present
time, we are unable to consider the energetics of quadruple-height
steps due to the size of this model \cite{limit}.

\begin{figure}
 \begin{center}
   \includegraphics[width=3.8in]{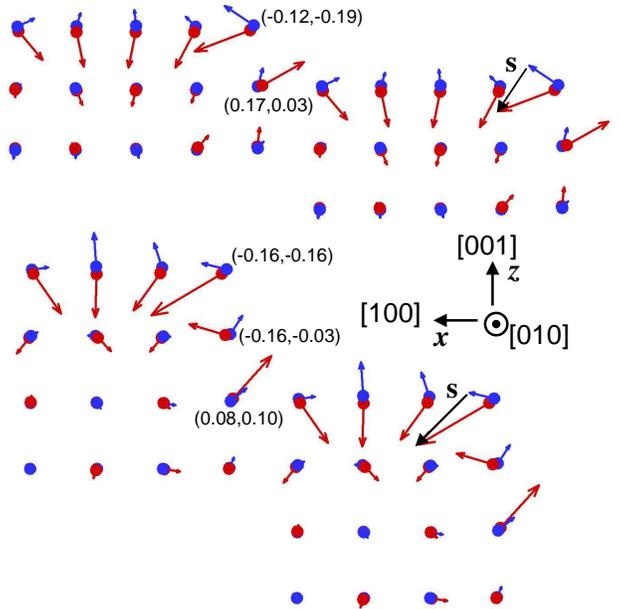}
 \end{center}
  \vspace{-4cm}
\caption{Atomic displacements of the TaC(105) and TaC(102)
surfaces with Sh steps (top) and Dh steps (bottom), respectively.
All vectors lie in the $xz$ plane and have been magnified for
clarity. Each atomic row is characterized by a rumpling vector,
whose components are given by the difference in displacements of
the Ta(red) and C(blue) atoms. The rumpling vector $\bf s$ of the
atomic row at the step-edge is marked in the figure. The
components of the rumpling vectors (in \AA) for a few atomic rows
near the step-edge are also given in the figure. }
\label{Dhdisplacements}
\end{figure}

We now turn our attention to the description of the steps in terms
of the atomic displacements at the step-edges, which are shown in
Fig.~\ref{Dhdisplacements} for the (105)-Sh and the (102)-Dh
surfaces. The figure shows that the atoms at the step-edges
undergo both horizontal ($x$) and vertical ($z$) displacements. On
moving a few atomic spacings away from the step-edge, the atoms on
the terraces are predominantly displaced in the vertical
direction, as in the case of the (001) surface. Overall, the
displacement field of the step resembles the field produced by a
force-dipole \cite{marchenko} located at the step-edge, though
there are new features that emerge. In particular, the step-edge
is rumpled, such that the horizontal displacements of the Ta atoms
at the step-edges are significantly larger than the corresponding
displacements of the C atoms. Since both the components of the
rumpling vector ${\bf s}$ given in Fig.~\ref{Dhdisplacements} are
comparable to the rumpling amplitudes for Ta(001) given in
Table~\ref{flatdispl}, we believe that this effect could be
observed in experiments. The step rumpling also leads to formation
of electric dipoles, which will be considered next.

\begin{figure}
  \begin{center}
  \includegraphics[width=3.5in]{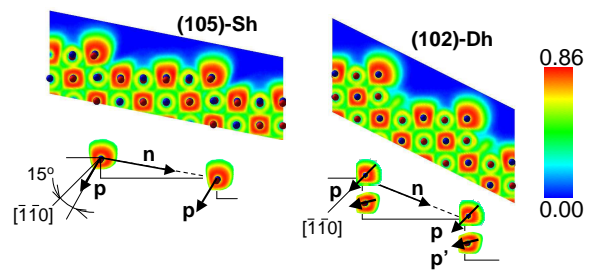}
  \vspace{-8.5cm}
\end{center}
\caption{Electron localization function (arbitrary units) in the
(010) cross-section for  TaC(105)-Sh and TaC(102)-Dh surfaces. The
dipoles of the C atoms at the step-edges are denoted by $\bf p$.
In the case of the Dh step, $\bf p'$ denotes the dipole of the C
atom located below the the Ta atom at the step-edge in the plane
immediately behind the one shown in the figure. } \label{elfcars}
\end{figure}

The magnitudes of the electric dipoles at the step-edges due to
rumpling can be estimated as  $Q|{\bf{s}}|$, where $Q$ is the
amount of charge transferred from C to Ta due to the partially
ionic character of the TaC-bond. Since $Q \sim |e|$ and
$|{\bf{s}}|\sim 0.1a$ (refer to Fig.~\ref{Dhdisplacements}), where
$e$ is the electronic charge, the linear dipole density due to
step rumpling is $P \sim 0.1|e|$. In addition to the dipoles that
arise as a result of step-rumpling, we also find that there are
additional dipoles that are created due to charge distortions at
the step-edge.  We observe that the valence shell around the C
atoms expands out of the surface leading to formation of electric
dipoles, denoted by ${\bf p}$ and ${\bf p}^{\prime}$ in
Fig.~\ref{elfcars}. Such electronic distortions also exist for the
Ta surface atoms, but to a much smaller extent. The magnitude of
these dipoles can be obtained from the electron localization
functions, which are shown in Fig.~\ref{elfcars} for the (105)-Sh
and the (102)-Dh surfaces. Our calculations show that the typical
magnitude the linear dipole density due to charge spill-out at the
step-edge is $P \sim 0.3-0.7|e|$.

The interaction energy between steps characterized by the
effective electrostatic dipole density $\bf P$ and the force
moment ${\bf D}=(D_x,D_z)$ is given by \cite{marchenko,Jay} :
\begin{equation}
\frac{\lambda_n^d}{L_x^2} =
\frac{\pi^2(P_{\perp}^2-P_{||}^2)}{3L_x^2} + \frac{\pi(1-\nu^2)
(\tau^2 n^2 a^2 + D_x^2)}{3 E L_x^2}, \label{U}
\end{equation}
where $P_{||}(\ P_{\perp}$) is the component of $\bf P$ parallel
(perpendicular) to the unit-vector $\bf n$ in Fig.~\ref{elfcars},
$\tau $ is the surface stress which generates the $z$-component of
the elastic dipole moment ($D_z=\tau n a$), $E$ is the Young's
modulus, and $\nu$ is the Poisson's ratio \cite{toth}. While
separating the elastic and electrostatic contributions to step
interactions is a difficult task, we can use the information from
Fig.~\ref{elfcars} to obtain estimates for these two effects.

First, observe that the dipole density due to step-rumpling is
about 20-30\% of the dipole density arising from charge spill-out;
it is therefore reasonable to calculate the electrostatic
interactions by focusing on the latter effect. Next, we see from
Fig.~\ref{elfcars} that compared to the Sh-step, the Dh-step has a
larger value of $P_{||}$ due to the presence of an additional C
atom with dipole ${\bf p}^{\prime}$ that is aligned closely with
the vector ${\bf n}$.  With increasing step-height, the number of
such atoms increases, leading to even larger values of $P_{||}$
and a  monotonic {\em decrease} in the electrostatic contribution
(refer to Eq.~(\ref{U})). However, we see from Table.~\ref{betas}
that the step interaction parameters show a non-monotonic trend,
which we attribute to an increase in the strength of elastic
interactions with the height of steps in the following paragraph.

 Using Eq.(\ref{U}), we calculate the electrostatic contribution
 to the interactions of Sh
 steps to be $\approx$ 1eV\AA , which is close to the value of $\lambda_1^d$ given in
 Table~\ref{betas}, indicating that the step interactions are dominated by
 the electrostatic effects. This observation is also in agreement with the
fact that a typical value of surface stress $\tau \approx$
100meV/\AA$^2$ yields a small repulsive elastic contribution which
is $\approx$ 0.01meV\AA. In the case of Dh-steps, a decrease in
step interaction parameter ($\lambda_2^d$) is consistent with a
decrease in the electrostatic contribution with increasing
step-height. However, a subsequent increase in the interaction
parameter ($\lambda_3^d$) for
 the Th-step can only be explained if the magnitude of the elastic dipole, $|D| = \sqrt{D_x^2 +
 D_z^2}$ increases with the height of the step. In this case, we expect
 both the electrostatic and elastic contributions to be comparable
 in magnitude.

In summary, by calculating step formation and interaction energies
we have shown that Dh and Th steps lead to more stable vicinal
surfaces, a result consistent with experimental observations
\cite{Zuo310}--\cite{Zuo910}. We have described the atomic
displacements of the surfaces and proposed a simple model for the
electrostatic interaction of the steps in terms of electric
dipoles of carbon atoms and step rumpling.

We gratefully acknowledge research support from the National
Science Foundation through grants CMS-0093714 and CMS-0210095 and
the Brown University MRSEC program. Computational support for this
work was provided by the National Center for Supercomputing
Applications through grant DMR-020032N and the Graduate School at
Brown University through the Salomon Research Award.


\begin{thebibliography}{99}

\bibitem{Zuo310}J.~K.~Zuo and D.~M.~Zehner, Phys. Rev. B {\bf 46}
16122 (1992).

\bibitem{Zuo110}J.~K.~Zuo, R.~J.~Warmack, D.~M.~Zehner and J.~F.~Wendelken,
Phys. Rev. B {\bf 47}, 10743 (1993).  

\bibitem{Zuostep}J.~K.~Zuo, J.~M.~Carpinelli, D.~M.~Zehner and J.~F.~Wendelken,
Phys. Rev. B {\bf 53}, 16013 (1996).  

\bibitem{Zuo910}J.~K.~Zuo, T.~Zhang, J.~F.~Wendelken and D.~M.~Zehner, Phys. Rev. B {\bf 63},
33404 (2001).  

\bibitem{Vasp}G.~Kresse and J.~Furthmuller, Phys. Rev. B {\bf 54}, 11169
(1996) and Comput. Mater. Sci. {\bf 6}, 15 (1996).

\bibitem{paw}G.~Kresse, and J.~Joubert, Phys. Rev. B {\bf 59}, 1758 (1999).

\bibitem{pw91}J.~Perdew and Y.~Wang, Phys. Rev. B {\bf 45}, 13244 (1992).

\bibitem{Gruzalski}G.~R.~Gruzalski, D.~M.~Zehner, J.~R.~Noonan, H.~L.~Davis,
R.~A.~Dio and K.~M\"{u}ller, J. Vac. Sci. Technol. A {\bf 7}, 3
(1989) and references therein.

\bibitem{TiC001}D.~L.~Price, J.~M.~Willis and B.~R.~Cooper, Phys. Rev.
Lett. {\bf 77} 3375 (1996) and Phys. Rev. B {\bf 48}, 15301
(1993).

\bibitem{Kobayashi}K.~Kobayashi, Jpn. J. Appl. Phys. {\bf 39}, 4311
(2000); K.~Kobayashi, Surf. Sci. {\bf 493}, 665 (2001).

\bibitem{Poon}T.~W.~Poon, S.~Yip, P.~S.~Ho and F.~F.~Abraham, Phys. Rev. B {\bf 46},
16122 (1992).

\bibitem{marchenko}V.~I.~Marchenko and Y.~A.~Parshin Sov. Phys. JETP
{\bf 52}, 129 (1980).

\bibitem{Jay}C.~Jayaprakash, C.~Rottmann, and W.~F.~Saam, Phys.
Rev. B {\bf 30}, 6549 (1984); A.~Redfield and A.~Zangwill Phys.
Rev. B {\bf 46}, 4289 (1992).

\bibitem{later}The fact that the point corresponding to $L_x=2.24$\AA \  lies {\em below} the
linear fit (Fig.~\ref{ledgefits}) indicates the presence a
short-range attraction between the steps, which has also been
inferred from experimental observations in Ref.~\cite{Zuo910}. We
defer a discussion on the origin of this short-range interaction
to a later publication.

\bibitem{einstein} B.~Jo\'{o}s, T.~L.~Einstein and N.~C.~Bartelt, Phys.
Rev. B {\bf 43}, 8153 (1991).

\bibitem{limit} Calculations of the energetics of quadrupole-height steps
become very demanding because, in addition to increased terrace
width, this system also requires larger slab and vacuum thickness
(more plane waves) in order to ensure that the interactions
involving the slab-faces and the slab-corners are small.

\bibitem{toth}The elastic constants for TaC are taken from
L.~E.~Toth, {\em Transition Metal Carbides and Nitrides} (Academic
Press, New York, 1971), page 148.




\end{thebibliography}
\end{document}